\documentclass[prl,aps,floatfix,superscriptaddress,showkeys,amsmath,amssymb,nofootinbib,twocolumn,10pt]{revtex4-1}
\usepackage{datetime}
\usepackage{xspace,array,tikz}
\usepackage{textcomp}
\usepackage{mathtools}
\usepackage{makeidx}
\usepackage{mathrsfs}
\usepackage{amsmath,graphicx}
\usetikzlibrary{shapes}

\newcommand{\DO}{{\hat{\rho}}}
\newcommand{\OpCalH}{\ensuremath{\hat {\mathcal{H}}}\xspace}

\newcommand{\Flux}{\ensuremath{\hat{\Phi}}}
\newcommand{\Charge}{\ensuremath{\hat{Q}}}

\newcommand{\COM}[2] {\ensuremath{\left[ {#1} , {#2}\right]}\xspace}
\newcommand{\ACOM}[2] {\ensuremath{\left\{ {#1} , {#2}\right\}}\xspace}

\newcommand{\Half} {\ensuremath{\frac{1}{2}}\xspace}

\newcommand{\ud}{\mathrm{d}} 
\newcommand{\ui}{\mathrm{i}} 

\DeclareMathOperator{\Tr}{Tr}

\newcommand{\be}{\begin{equation}}
\newcommand{\bel}[1]{\begin{equation}\label{#1}}
\newcommand{\ee}{\end{equation}}
\newcommand{\ba}{\begin{eqnarray}}
\newcommand{\ea}{\end{eqnarray}}
\newcommand{\bal}{\begin{align}}
\newcommand{\eal}{\end{align}}

\newcommand{\TD}[2]{\ensuremath{\frac{\ud #1}{\ud #2}}\xspace}

\newcommand{\Fig}[1]{Fig.~\ref{#1}\xspace}
\newcommand{\Eq}[1]{Eq.~(\ref{#1})\xspace}

\newcommand{\Opp}{\ensuremath{\hat p}\xspace}
\newcommand{\Opx}{\ensuremath{\hat x}\xspace}
\newcommand{\OpP}{\ensuremath{\hat P}\xspace}
\newcommand{\OpA}{\ensuremath{\hat A}\xspace}

\newcommand{\OpL}{\ensuremath{\hat {L}}\xspace}

\newcommand{\OpX}{\ensuremath{\hat X}\xspace}

\newcommand{\OpH}{\ensuremath{\hat H}\xspace}

\newcommand{\QSERG}{Quantum Systems Engineering Research Group, Loughborough University, Loughborough, Leicestershire LE11 3TU, United Kingdom}
\usepackage{amssymb,amsmath,mathrsfs,dsfont,color}
\newcommand{\LBORO}{Department of Physics, Loughborough University}
\begin{document}
\title{On open quantum systems, effective Hamiltonians and device characterization}

\author{S.N.A.~Duffus}
\affiliation{\QSERG}
\affiliation{\LBORO}
\author{V.M.~Dwyer}
\email{v.m.dwyer@lboro.ac.uk}
\affiliation{\QSERG}
\affiliation{The Wolfson School of Mechanical, Electriical and Manufacturing Engineering, Loughborough University}
\author{M.J.~Everitt}
\affiliation{\QSERG}
\affiliation{\LBORO}

\begin{abstract}
High fidelity models, which support accurate device characterization and correctly account for environmental effects, are crucial to the engineering of scalable quantum technologies. As it ensures positivity of the density matrix, one preferred model for open systems describes the dynamics with a master equation in Lindblad form. The Linblad operators are rarely derived from first principles, resulting in dynamical models which miss those additional terms that must generally be added to bring the master equation into Lindblad form, together with concomitant other terms that must be assimilated into an effective Hamiltonian. In first principles derivations such additional terms are often canceled (countered), frequently in an \textit{ad hoc} manner. In the case of a Superconducting Quantum Interference Device (SQUID) coupled to an Ohmic bath, the resulting master equation implies the environment has a significant impact on the system's energy. We discuss the prospect of keeping or canceling this impact; and note that, for the SQUID, measuring the magnetic susceptibility under control of the capacitive coupling strength and the externally applied flux, results in experimentally measurable differences between models. If this is not done correctly, device characterization will be prone to systemic errors. 
\end{abstract}

\maketitle

For any complex system, if device characterization and design engineering are to be meaningful notions, it is necessary to have access to device models which faithfully describe behavior and performance when subjected to a wide range of environmental conditions. The ability to integrate billions of CMOS transistors onto a single chip relies on careful device characterization as well as knowledge of likely environmental conditions. A chip meant for liquid Helium applications will be characterized by lower junction capacitances, as a result of carrier freeze-out \cite{lowT}, whilst if intended for space applications, it must be designed with radiation hardness in mind \cite{radiationhardened}. Successful Quantum Technologies will be no different in this regard. 
In the quest to build a scalable quantum processor from superconducting circuits containing Josephson junctions, significant developments have been made recently; including enhanced coherence times \cite{coherencetime}, high fidelity state-preparation and measurement \cite{HiFistateprep}, and the demonstration of two-qubit gates \cite{Qubitgates}. To facilitate the fabrication of functional circuits, future qubits must be capable of long and short term storage; be frequency tunable; be capable of communication over both short and long ranges, and possess tunable couplings \cite{1367-2630-11-3-033030}. Decoherence presents a formidable obstacle to achieving this for superconducting qubits. As a result, the maintenance of a long coherence time has become a key problem in superconductor qubit research; making the need for a better understanding of different types of environment a similarly key concern.  

Until relatively recently it had been assumed that the connection to ground of floating flux-qubits was poor and that such qubits were immune to capacitive coupling. However a recent study \cite{1367-2630-11-3-033030} has demonstrated the reactance to ground can become sizeable, and that not protecting against this decoherence channel may explain some of the short coherence times obtained previously. The equivalent design paradigm for a classical semiconductor (CMOS) computer allows a very detailed simulation of critical-path devices in a variety of environments (at fast/typical/slow mobility process corners; under mechanical stress, flicker or thermal noise, etc.) as well in a range of operational conditions (IR-drop on power lines) and including a number of reliability considerations (metal migration) before it is committed to silicon \cite{MOSFET}. In the particular case of capacitive coupling, the contrast of the superconducting qubit to silicon CMOS is stark; indeed CMOS is the chosen platform because of its excellent noise immunity. Understanding the impact of an environment is necessary for accurate characterization of a given quantum device; as devices which are not characterized under the same environmental conditions may not be easily compared, and a device characterized in a test environment, may behave differently in use.

A common means of modeling an open quantum system is with a reduced master equation~\cite{PhysRevLett.110.086403,breuer2007theory,
PhysRevLett.58.792,
PhysRevLett.108.040401,
PhysRevLett.102.080405,
PhysRevLett.116.120402,
PhysRevA.89.042120,
PhysRevA.86.042107,
PhysRevA.90.063815,
PhysRevLett.117.230401,
PhysRevLett.113.200403}, 
as these approximate the dynamics of the system of interest without the need to model the environment itself, and, because it guarantees certain nice properties (such as conservation of probability and complete positivity of the reduced (system) density matrix), it is usual to cast the master equation into Lindblad form~\cite{LINDBLAD1976393}. In this model Lindblad operators represent the action of the environment on the system. Exactly which operators to use usually involves a careful construction analysis of the system + environment, followed by a tracing out of the environmental degrees of freedom. This process requires a number of approximations, as well as some decisions about the form of the final equation. 
Most common is the Born-Markov (BM) approximation which assumes a weak coupling between a quantum system and its environment, assumed to have no long-term memory. However the BM approximation alone will not generally yield a quantum master equation in Lindblad form. 
As a consequence, it is common to manipulate the resulting BM master equation by adding those terms necessary to complete the Lindblad structure and by either absorbing any terms which present as free evolution into an effective system Hamiltonian, or by canceling them with heuristic counter terms. 

The simplest example is that of Quantum Brownian Motion (QBM), with a Harmonic Oscillator (HO) potential, coupled to an Ohmic bath through the system's position operator $\Opx$, e.g. \cite{breuer2007theory}.
Here a Lindblad master equation can be created from Born-Markov development by: (i) adding a term of the form $D_{pp} [\Opp, [\Opp, \hat \rho]]$, where $\Opp$ is the canonical conjugate momentum to \Opx ($\COM{\Opx}{\Opp}=\ui\hbar$), and $D_{pp}$ is inversely proportional to the bath temperature $T$; and (ii) by moving terms proportional to $[\Opx^2,\hat \rho ]$ (the Lamb shift term) and $\mu[\Opx\Opp+\Opx\Opp, \hat \rho]$ (a squeezing term, where $\mu$ depends on system-bath coupling strength) into an effective system Hamiltonian. 
The justification for the addition of the $D_{pp}[\Opp, [\Opp, \hat \rho]]$ term is often that, in the high $T$ limit, it will is small and so amounts to a minimally invasive means of ensuring Lindblad form.  
The Lamb shift term $[\Opx^2,\hat \rho ]$ is generally canceled by the inclusion of an identical (counter) term in the system or interaction Hamiltonian on the grounds that it otherwise constitutes an unphysical frequency renormalization.

 Up to this point what has been described is the standard high-temperature (Lindblad) version of the Caldiera-Leggett equation~\cite{CALDEIRA1983374}, who argued that, for many systems, a counter term arises to cancel the Lamb shift in a truly detailed analysis of microscopic dynamics.  
The final (squeezing term) $\mu [\Opx\Opp+\Opp\Opx, \hat \rho]$, has been discussed in the literature on a number of occasions. On the one hand its presence (uncountered) in an effective Hamiltonian is necessary to ensure: translational invariance~\cite{PhysRevLett.79.3101}; that Erhenfest's theorem is satisfied~\cite{LINDBLAD1976393,SANDULESCU1987277,PhysRevLett.79.3101,PhysRevLett.80.5702,PhysRevA.94.042123}; and a correct quantum to classical transition
~\cite{
0305-4470-28-18-028,
1367-2630-11-1-013014, 
Everitt20102809,
1742-6596-306-1-012045,
PhysRevA.79.032328}. 

  On the other hand, the inclusion of a counter term $-\lambda [\Opx\Opp+\Opp\Opx, \hat \rho]$ in the system Hamiltonian has occasionally been part of an ansatz which assumes the most general second order Lindblad master equation possible, using a single (annihilator) Lindblad, first order in system variables $\Opx$ and $\Opp$, together with a general second order Hamiltonian. Its properties have been investigated as a function of $\lambda$. Setting $\lambda=\mu$ cancels the squeezing term completely, while $\lambda \neq 0$ gives additional flexibility to add desirable properties to the master equation, such as reasonable low $T$ behavior~\cite{PhysRevLett.79.3101}. 
This does however sacrifice translational invariance~\cite{LINDBLAD1976393}, and Ehrenfest~\cite{PhysRevLett.80.5702}. 

 The difficulties with the derivation of the Lindblads appropriate to a particular system-environment interaction are clearer still when considering a particle in an anharmonic potential, such as is the case for a Superconducting Quantum Interference Device (SQUID). Here the heuristic adding/canceling of terms becomes more involved~\cite{PhysRevB.94.064518}, which necessarily raises difficult issues, as high-precision control of such a system inevitably requires precise device characterization~\cite{PhysRevA.71.062312}. 

We consider the following standard model of a SQUID (charge $\Charge$ and flux $\Flux$, $[\Flux,\Charge]=\ui\hbar$), inductively and capacitively coupled to an Ohmic bath environment, modeled as an infinite set of harmonic oscillators, indexed by $n$, with charge $\Charge_n$ and flux $\Flux_n$ ($[\Flux_n,\Charge_m]=\ui\hbar \delta_{nm}$). The dimensionless total Hamiltonian $\OpH/\hbar\omega_0$ may be written as a sum of the Hamiltonians of the SQUID ($\OpCalH_S$), the bath ($\OpCalH_B$), the coupling between them $(\OpCalH_I)$  
\vspace*{-4pt}
 \ba \label{eq:HAM0} 
 \OpCalH_{S}&=&\frac{\OpX^{2}}{2}+\frac{\OpP^{2}}{2}-\frac{\nu}{\omega_0}\cos \left(\sqrt{\frac{\beta\nu}{\omega_0}}\OpX+ \frac{2\pi\Phi_x}{\Phi_0}\right) + \OpCalH_{LS} \nonumber\\ 
\OpCalH_{B}&=& \sum_{n}\frac{\OpX^{2}_{n}}{2}+\frac{\OpP_n^{2}}{2},\ \   
 \OpCalH_I =-\sum_n \kappa_n\left( \OpX \OpX_n + g\OpP \OpP_n \right)  
\ea 
where $L$ and $C$ are the system's inductance and capacitance, $\omega_0=1/\sqrt{LC}$, $\OpP=\Charge(\hbar C \omega_0)^{-1/2}$, $\beta\nu=4\pi^2\hbar/\Phi_0^2 C$, $\OpP_n=\Charge_n(\hbar C_n \omega_0)^{-1/2}$, $\OpX=\Flux(\hbar L \omega_0)^{-1/2}$, and $\OpX_n=\Flux_n(\hbar L_n \omega_0)^{-1/2}$, while $\Phi_0$ denotes the flux quantum. The externally applied flux, $\Phi_x$, controls the phase of the cosine in \Eq{eq:HAM0} which describes the coupling across the Josephson junction (energy $=\hbar\nu$) \cite{0034-4885-76-7-076001}. The strengths of the environmental couplings are determined by $\kappa_n$ and $g$, the ratio of inductive to capacitive coupling. $\OpCalH_{LS}$ is the Lamb Shift Hamiltonian which will be canceled in the Lindbald process~\cite{CALDEIRA1983374}.

In the interaction picture, the BM master equation is
\setlength{\abovedisplayskip}{0pt} 
\setlength{\abovedisplayskip}{0pt} 
 \bel{eq:genME} \begin{split}  
	&\TD{\DO_{S}(t')}{t'}= \mathcal{L}[\DO]+\mathcal{K}[\DO]\\ 
    &=-\ui[\OpCalH_{S},\DO_{S}(t')] -\int_0^\infty \ud\tau\Tr_{B} \left[\OpCalH_{I},\left[\OpCalH_{I}(-\tau),\DO_{S}(t')\otimes \DO_{B}\right]\right] \nonumber
 \end{split}\ee
\cite{breuer2007theory}.The first term describes free evolution and the second, the dissipator, describes non-unitary loss. 
We assume an Ohmic spectral function
$
J(\omega)=2C\gamma \omega \Omega^2/\pi(\Omega^2+\omega^2)
$ 
with damping rate $\gamma$ and bath cut-off frequency $\Omega$. 
Substituting $\OpCalH_{I}$, from~\Eq{eq:HAM0}, into the dissipator integral and evaluating the bath correlation functions yields \cite{breuer2007theory}
\begin{widetext}
\ba 
\mathcal{K}[\DO] &=  &\frac{\gamma \Omega}{2\omega_0}\int_{0}^{\infty}\ud\tau e^{-\Omega\tau}\Bigg(  \frac{2\Omega}{\omega_0} \bigg( \ui\COM{\OpX}{\ACOM{\OpX(-\tau)}{\DO_{S}}}+\ui g^2\COM{\OpP}{\ACOM{\OpP(-\tau)}{\DO_S}}
+g \COM{\OpX}{\COM{\OpP(-\tau)}{\DO_S}}- g \COM{\OpP}{\COM{\OpX(-\tau)}{\DO_S}}\bigg) \nonumber \\
 &&- \bigg(\COM{\OpX}{\COM{\OpX(-\tau)}{\DO_{S}}}  + g^2 \COM{\OpP}{\COM{\OpP(-\tau)}{\DO_S}}
-\ui g \COM{\OpX}{\ACOM{\OpP(-\tau)}{\DO_S}} 
+ \ui g \COM{\OpP}{\ACOM{\OpX(-\tau)}{\DO_S}}\bigg) \Bigg)
\ea 
\noindent where $\ACOM{\cdot}{\cdot}$ denotes anti-commutation.
Then, evaluating the integrals, and expanding the correlation time-dependent observables $\OpX(-\tau)$ and $\OpP(-\tau)$ to first order, yields the master equation:
\ba \label{eq:ndd}
m\TD{\DO_{S}}{t'} &=&-\ui\COM{\OpCalH_{S}}{\DO_{S}}
-\frac{\ui\gamma}{\omega_0}\left(1+g^2-g\right) \COM{\OpX}{\ACOM{\OpP}{\DO_S}} 
-\frac{\ui\gamma}{\omega_0}g\left(g-\Half\right) \COM{\ACOM{\OpX}{\OpP}}{\DO_S}\\ \nonumber
&&-\frac{\gamma}{\omega_0}\left(g+\frac{1}{2}\right)\COM{\OpX}{\COM{\OpX}{\DO_S}} 
+ \frac{\gamma}{2\Omega}\left(1-g^2\right)\COM{\OpX}{\COM{\OpP}{\DO_S}} 
-\frac{\gamma}{\omega_0}g\left(1+\frac{g}{2}\right)\COM{\OpP}{\COM{\OpP}{\DO_S}}\\ \nonumber
&&+\frac{\ui \gamma g}{\omega_0}\sqrt{\frac{\beta\nu}{\omega_0}} \COM{\frac{\omega_0}{2\Omega}\OpX +g\OpP}{\ACOM{\sin{\left(\sqrt{\frac{\beta \omega_0}{\nu}}\OpX +2\pi\frac{\Phi_x}{\Phi_0}\right)}}{\DO_S}}
- \frac{\gamma g}{\omega_0} \sqrt{\frac{\beta\nu}{\omega_0}}\COM{\OpX +\frac{g \omega_0}{2\Omega}\OpP}{\COM{\sin{\left(\sqrt{\frac{\beta \omega_0}{\nu}}\OpX +2\pi\frac{\Phi_x}{\Phi_0}\right)}}{\DO_S}}
%
\ea
\end{widetext}
In arriving at~\Eq{eq:ndd} two quadratic frequency renormalization terms (one proportional to $[\OpX^2,\DO]$ and one proportional to $[\OpP^2,\DO]$) have been removed, canceled by the Lamb Shift term $\OpCalH_{LS}$~\cite{PhysRevE.49.1854,Spiller1991}. Note also that the external flux control $\Phi_x$ appears in the dissipator in this first order model (as opposed to second order for the case of flux-only coupling 
\cite{0034-4885-76-7-076001}). 
 ~\Eq{eq:ndd} is a non-rotating wave (NRW) equation, and so requires additional terms to assume Lindblad form \cite{Munro1996}
\bel{eq:Lindblad}
\frac{\ud\DO}{\ud t} = -\ui[\OpCalH',\DO] + \frac{1}{2}\sum_{j}\bigg\{[\OpL_{j}, \DO \OpL_{j}^{\dagger}] + [\OpL_{j}\DO, \OpL_{j}^{\dagger}]\bigg\}
\ee
where $\OpL_j$ are Lindblad operators. From~\Eq{eq:ndd} it is clear that these will be linear combinations of $\OpX$, $\OpP$, and $\hat S=\sin{\left(\sqrt{\frac{\beta\omega_0}{\nu}}\OpX+ 2\pi\frac{\Phi_x}{\Phi_0}\right)}$, whose weights may be obtained by diagonalizing the coefficient matrix \cite{breuer2007theory}
\vspace*{-4pt}
\begin{widetext}
\bel{eq:coefmat}
a_{ij}= \frac{\gamma}{\omega_0}\begin{pmatrix} 2g+1 & -\ui (1+g^2-g) - \xi (1-g^2) & g \sqrt{\frac{\beta\nu}{\omega_0}} \left(1+\xi\right)\\
 \ui (1+g^2-g) -\xi (1-g^2)& 2g+g^2& g^2\sqrt{\frac{\beta\nu}{\omega_0}}\left(\xi+\ui\right)
  \\ g \sqrt{\frac{\beta\nu}{\omega_0}} \left(1-\ui\xi\right)&g^2\sqrt{\frac{\beta\nu}{\omega_0}}\left(\xi-\ui\right) & 0\end{pmatrix}
\ee
\end{widetext}
\vspace*{-12pt}
for $i,j \in \{X,P,S\}$ and $\xi=\omega_0/2\Omega$. As is often found with NRW BM equations, the coefficient matrix is not positive semidefinite, and so does not conserve probability. To ensure physicality we alter $a_{SS}$ just sufficient to make the $a$ matrix positive semi-definite, and its eigenvalues non-negative. Effectively this is accomplished in a \textit{minimally invasive} \cite{breuer2007theory} manner by setting $\det(a) = 0$, i.e. 
\bel{eq:coefmat33}
a_{SS}=\frac{\beta \nu}{\omega_0}\frac{-\xi^2g^5+4g^4+8\xi^2g^3}{(-g^4+2g^2-1)(1+\xi^2) +4g(g^2+1)}
\ee
together with the constraint that $0.227 \le g \le 4.40$. This procedure ensures the two non-zero eigenvalues are positive and, whilst the range of $g$ values may be extended, it does come at the cost of a more invasive change to the matrix $a$. For our present purposes the details of the Lindblads are not particularly important; though we note that the dominant one is close to an annihilator, of $c_1 \OpX + c_2\OpP$. Instead, here we are interested in the question of what one might do with the terms that are leftover in the process. As in the case of the QBM with a harmonic potential, all these terms present in the form of a unitary evolution (i.e. as $\ui[\OpA, \hat \rho]$), however there appears to be no clear means of assessing whether they should be kept, to produce an effective Hamiltonian, or canceled (countered) - either partially or completely. Retaining all such terms for now, we obtain an effective Hamiltonian $\OpCalH'= \OpCalH_S + \OpCalH_{XP} + \OpCalH_{PS} + \OpCalH_{XS}$ where
\bel{eq:HAM}\begin{split}
\mathcal{\OpH}_{XP} &=\frac{\gamma}{2\omega_0} \left(\frac{3g^2}{2}-g+\Half\right)(\OpX\OpP+\OpP\OpX),\\
\mathcal{\OpH}_{XS}&= -\frac{\gamma g}{2\Omega}\sqrt{\frac{\beta\nu}{\omega_0}}\OpX\sin{\left(\sqrt{\frac{\beta \omega_0}{\nu}}\OpX +2\pi\frac{\Phi_x}{\Phi_0}\right)}, \\ 
\mathcal{\OpH}_{PS}&= -\frac{\gamma g^2}{2\omega_0}\sqrt{\frac{\beta\nu}{\omega_0}}\ACOM{\OpP}{\sin{\left(\sqrt{\frac{\beta \omega_0}{\nu}}\OpX 2\pi\frac{\Phi_x}{\Phi_0}\right)}}\\
%
%
&-\frac{\gamma g^2\beta}{4\Omega} \cos{\left(\sqrt{\frac{\beta \omega_0}{\nu}}\OpX +2\pi\frac{\Phi_x}{\Phi_0}\right)}.
\end{split}\ee

\begin{figure}[!tb]
\centering
\includegraphics[width=\linewidth]{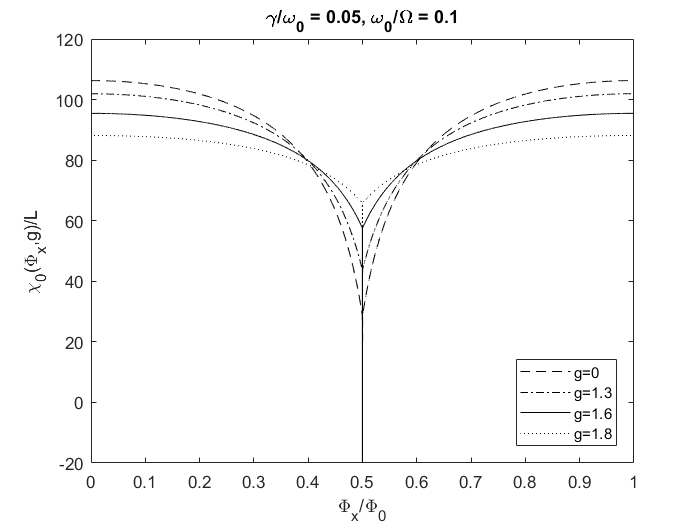}
\setlength{\abovecaptionskip}{-5pt}
\setlength{\belowcaptionskip}{-10pt}
\caption{Ground state magnetic susceptibility as a function of the external flux $\Phi_x$, and capacitive coupling strengths $g$.\\
}\label{fig:suscept}

\end{figure}

 \newcommand{\D}{8} 
  \newcommand{\U}{60} 
  \newdimen\R 
  \R=3.5cm  
  \newdimen\L 
  \L=4.0cm
  \newcommand{\A}{360/\D} 
The impact of $\OpCalH_{XP}$ has previously been discussed in terms of frequency shifts and squeezing \cite{0034-4885-76-7-076001,PhysRevB.94.064518} and, as previously noted, in other contexts too~\cite{PhysRevLett.79.3101,LINDBLAD1976393,SANDULESCU1987277,PhysRevLett.79.3101,PhysRevLett.80.5702,PhysRevA.94.042123,0305-4470-28-18-028,1367-2630-11-1-013014, 
PhysRevA.79.032328},
whilst a term similar in nature to $\OpCalH_{XS}$ appears in the higher order (in $\omega_0/\Omega$) master equation for a purely inductively coupled SQUID \cite{0034-4885-76-7-076001}. The final term $\OpCalH_{PS}$ includes a slight renormalization of the Josephson junction energy. 
  \begin{figure}[!tb]
  \resizebox{0.9\linewidth}{!}{
  \includegraphics[trim={3cm 7cm 3cm 8cm},width=\linewidth]{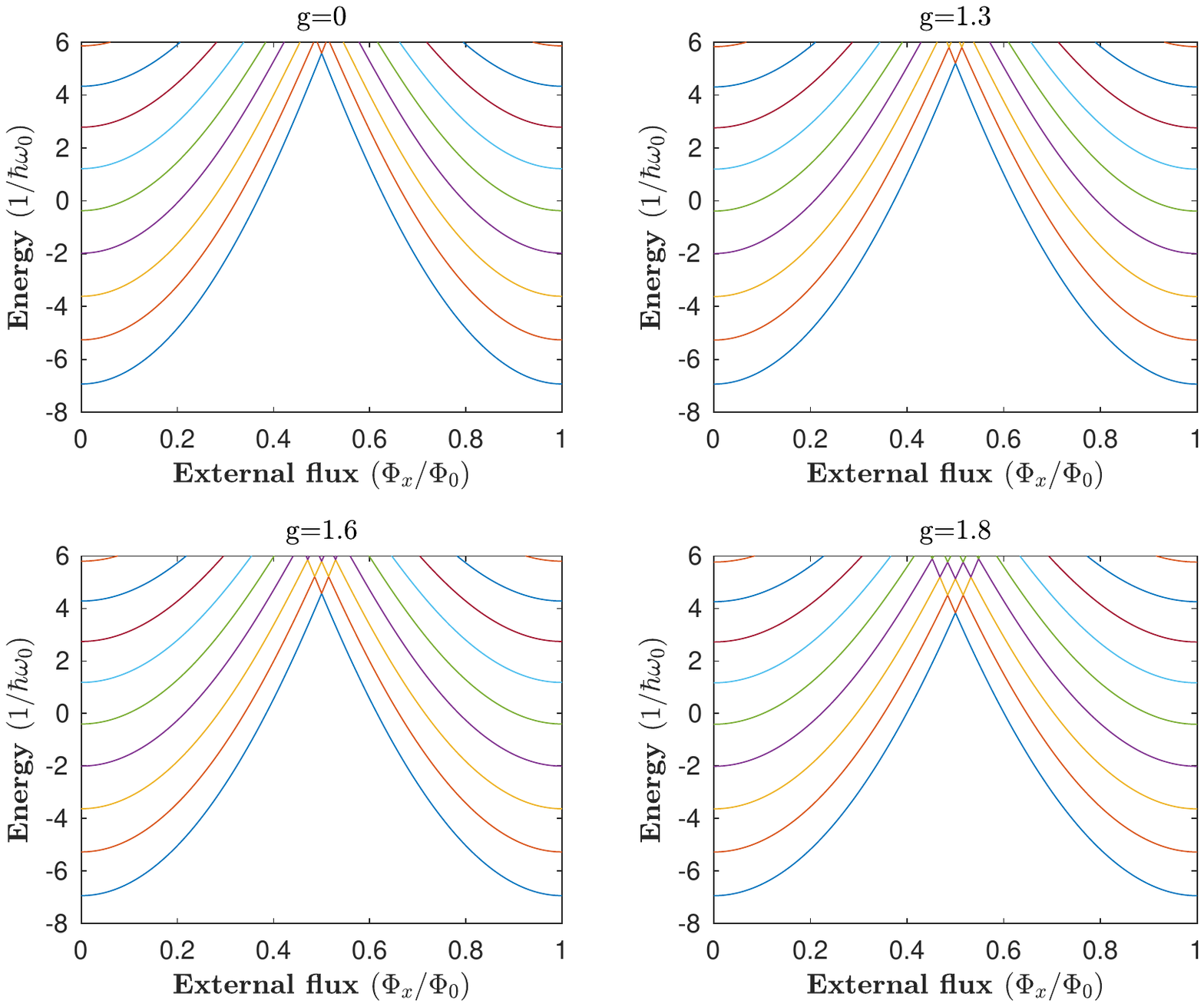}}
  \caption{The energy eigenvalues of $\OpCalH'$ as a function of external flux $\Phi_x$ for various coupling ratios $g$. Values obtained with Josephson energy $\hbar\nu = 6.693\times 10^{-22}$J,capacitance $C_0=5\times 10^{-15}$F, inductance $L_0= 3\times 10^{-10}$H, damping rate $\gamma = 0.05\omega_0$ and bath frequency $\Omega = 10*\omega_0$.\label{fig:evals}
  }
  \end{figure}
~\Fig{fig:evals} demonstrates the impact of these terms on the lowest five energy eigenvalues of $\OpCalH'$ as a function of the external flux $\Phi_x$, and for a variety of $g$ values.
\begin{figure}[!tb]
 \centering
 \resizebox{\linewidth}{!}{
 \begin{tikzpicture}[scale=1]
  \path (0:0cm) coordinate (O); 

  \foreach \X in {1,...,\D}{
    \draw (\X*\A:0) -- (\X*\A:\R);
  }

  \foreach \Y in {0,...,\U}{
    \foreach \X in {1,...,\D}{
      \path (\X*\A:\Y*\R/\U) coordinate (D\X-\Y);
      \fill (D\X-\Y) circle (0.5pt);
    }
    \draw [opacity=0.00] (0:\Y*\R/\U) \foreach \X in {1,...,\D}  {
        -- (\X*\A:\Y*\R/\U)
    } -- cycle;
  }
  \path (1*\A:\L) node (L1) {\hspace{60pt} $\OpCalH'= \OpCalH_S + \OpCalH_{XP} + \OpCalH_{PS} + \OpCalH_{XS}$};
  \path (2*\A:\L) node (L2) { $\OpCalH_0$};
  \path (3*\A:\L) node (L3) { $\OpCalH_0 +\OpCalH_{XS}$};
  \path (4*\A:\L) node (L4) [anchor=east]{ $\OpCalH_0+ \OpCalH_{XP}$};
  \path (5*\A:\L) node (L5)[anchor=north east] { $\OpCalH_0+\OpCalH_{PS}$};
  \path (6*\A:\L) node (L6) { $\OpCalH_0+\OpCalH_{XP}+\OpCalH_{XS}$};
  \path (7*\A:\L) node (L7)[anchor=north] { $\OpCalH_0+\OpCalH_{XP}+\OpCalH_{PS}$};
  \path (8*\A:\L) node (L8) [anchor=west]{ $\OpCalH_0+\OpCalH_{PS}+\OpCalH_{XS}$};
  \draw [color=red,line width=1.5pt,opacity=0.5]
   (D1-38) --
    (D2-56) --
    (D3-57) --
    (D4-47) --
    (D5-54) -- 
    (D6-48)--
    (D7-38)--
    (D8-55)--cycle;
  \draw [color=blue,style=dashed,line width=1pt,opacity=0.5]
    (D1-56) --
    (D2-56) --
    (D3-56) --
    (D4-56) --
    (D5-56) -- 
    (D6-56) --
    (D7-56) --
    (D8-56) --cycle;
  \draw [color=blue,style=dotted,line width=1pt,opacity=0.5]
    (D1-38) --
    (D2-38) --
    (D3-38) --
    (D4-38) --
    (D5-38) -- 
    (D6-38) --
    (D7-38) --
    (D8-38)--cycle;
  \end{tikzpicture}}
  \caption{Spiderweb Diagram. The solid line shows the lowest eigenvalue for the SQUID Hamiltonian plus combinations of additional terms as labeled. Dashed and dotted lines show, for comparison, the lowest eigenvalues of $\OpCalH_0$ and $\OpCalH'$ respectively. Note that the contributions of $\OpCalH_{XP}$,$\OpCalH_{XS}$ and $\OpCalH_{PS}$ are not independent, and the combination $\OpCalH_{XP} +\OpCalH_{XS}$ accounts for the majority of the alteration. The origin is $E=0$, the outer radius is $E=6.0\hbar\omega_0$. Here $\Phi_x=0.5\Phi_0$, $g=1.8$.}
  \label{fig:spiderweb}
\end{figure}
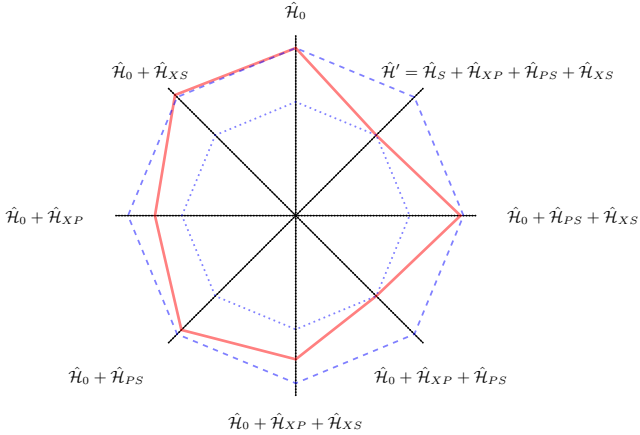
\Fig{fig:spiderweb} shows the contribution of $\OpCalH_{XP}, \OpCalH_{PS}$ and $\OpCalH_{XP}$ to the lowest energy level, both individually and in pairs, and illustrates how the cancellation, or not, of such terms will alter the system energy eigenstates. It is clear that these effects are non-negligible, and that any form of \textit{ad hoc} process to deal with their contribution is likely to lead to erroneous system models. That one of the terms is known to be important in the quantum to classical transition~\cite{
0305-4470-28-18-028,
1367-2630-11-1-013014, 
Everitt20102809,
1742-6596-306-1-012045,
PhysRevA.79.032328} implies that there are physically important implications here which probably should not be ignored.
The magnetic susceptibility for a SQUID ring 

\be
\chi_0(\Phi_x;g) = - L \frac{\partial^2 E_0 (\Phi_x; g)}{\partial \Phi_x^2},
\ee
in its ground state, provides a useful mechanism to probe the lowest energy eigenvalue's dependence on the external flux $\Phi_x$ and coupling ratio $g$ \cite{SQUID}. The ground state magnetic susceptibility (in the form $\chi_0(\Phi_x;g)/L$) is shown in~\Fig{fig:suscept} where it is seen to vary significantly with both $\Phi_x$ and $g$. 
An important conclusion is that, by arranging an adjustable capacitive coupling, and varying the external flux, a measurement of $\chi_0$ provides an empirical test of the presence of these terms; and hence whether they are indeed physical and need to be kept, or merely an artifact of the process needed to shoehorn Born-Markov master equations into Lindblad form, and need to be canceled.      

 Making new Quantum Technologies scalable will rely on the ability to perform careful modeling and simulation. Any simulation framework will require high fidelity models and accurate characterization methods to allow precision control and enable the necessary error correction. However it seems clear that, beyond the simple case of a harmonic potential, deriving Lindblad master equations from the Born-Markov approximation becomes rather \textit{ad hoc}, as the decision to include (as being physically real) or to exclude (through the use of counter terms) is not supported well enough by our knowledge of the systems involved. Even in that simplest case (QBM) there is still some disagreement about whether a squeezing term of the form $\OpX\OpP+\OpP\OpX$ should be kept or canceled \cite{SANDULESCU1987277,PhysRevLett.79.3101,PhysRevLett.80.5702,PhysRevA.94.042123}. 
The SQUID system considered here has simple, adjustable control parameters: the external flux $\Phi_x$ and the relative strength $g$ of capacitive coupling. Together they appear to provide a sufficient means of interrogating the energy level structure through the magnetic susceptibility. In the model considered, the additional terms will make a measurable difference. Experimentally, the ability to create superconducting devices coupled to an artificially constructed bath of harmonic oscillators is well within the state-of-the-art. Hence, it will be possible to experimentally verify whether, for example the squeezing term $\OpX\OpP+\OpP\OpX$ should be canceled by a counter term.
\bibliographystyle{apsrev4-1}
\bibliography{REFCC}

\begin{thebibliography}{36}%
\makeatletter
\providecommand \@ifxundefined [1]{%
 \@ifx{#1\undefined}
}%
\providecommand \@ifnum [1]{%
 \ifnum #1\expandafter \@firstoftwo
 \else \expandafter \@secondoftwo
 \fi
}%
\providecommand \@ifx [1]{%
 \ifx #1\expandafter \@firstoftwo
 \else \expandafter \@secondoftwo
 \fi
}%
\providecommand \natexlab [1]{#1}%
\providecommand \enquote  [1]{``#1''}%
\providecommand \bibnamefont  [1]{#1}%
\providecommand \bibfnamefont [1]{#1}%
\providecommand \citenamefont [1]{#1}%
\providecommand \href@noop [0]{\@secondoftwo}%
\providecommand \href [0]{\begingroup \@sanitize@url \@href}%
\providecommand \@href[1]{\@@startlink{#1}\@@href}%
\providecommand \@@href[1]{\endgroup#1\@@endlink}%
\providecommand \@sanitize@url [0]{\catcode `\\12\catcode `\$12\catcode
  `\&12\catcode `\#12\catcode `\^12\catcode `\_12\catcode `\%12\relax}%
\providecommand \@@startlink[1]{}%
\providecommand \@@endlink[0]{}%
\providecommand \url  [0]{\begingroup\@sanitize@url \@url }%
\providecommand \@url [1]{\endgroup\@href {#1}{\urlprefix }}%
\providecommand \urlprefix  [0]{URL }%
\providecommand \Eprint [0]{\href }%
\providecommand \doibase [0]{http://dx.doi.org/}%
\providecommand \selectlanguage [0]{\@gobble}%
\providecommand \bibinfo  [0]{\@secondoftwo}%
\providecommand \bibfield  [0]{\@secondoftwo}%
\providecommand \translation [1]{[#1]}%
\providecommand \BibitemOpen [0]{}%
\providecommand \bibitemStop [0]{}%
\providecommand \bibitemNoStop [0]{.\EOS\space}%
\providecommand \EOS [0]{\spacefactor3000\relax}%
\providecommand \BibitemShut  [1]{\csname bibitem#1\endcsname}%
\let\auto@bib@innerbib\@empty
\bibitem [{\citenamefont {Yoshikawa}\ \emph {et~al.}(2005)\citenamefont
  {Yoshikawa}, \citenamefont {Tomida}, \citenamefont {Tokuda}, \citenamefont
  {Liu}, \citenamefont {Meng}, \citenamefont {Whiteley},\ and\ \citenamefont
  {Duzer}}]{lowT}%
  \BibitemOpen
  \bibfield  {author} {\bibinfo {author} {\bibfnamefont {N.}~\bibnamefont
  {Yoshikawa}}, \bibinfo {author} {\bibfnamefont {T.}~\bibnamefont {Tomida}},
  \bibinfo {author} {\bibfnamefont {M.}~\bibnamefont {Tokuda}}, \bibinfo
  {author} {\bibfnamefont {Q.}~\bibnamefont {Liu}}, \bibinfo {author}
  {\bibfnamefont {X.}~\bibnamefont {Meng}}, \bibinfo {author} {\bibfnamefont
  {S.~R.}\ \bibnamefont {Whiteley}}, \ and\ \bibinfo {author} {\bibfnamefont
  {T.~V.}\ \bibnamefont {Duzer}},\ }\href {\doibase 10.1109/TASC.2005.849786}
  {\bibfield  {journal} {\bibinfo  {journal} {IEEE Transactions on Applied
  Superconductivity}\ }\textbf {\bibinfo {volume} {15}},\ \bibinfo {pages}
  {267} (\bibinfo {year} {2005})}\BibitemShut {NoStop}%
\bibitem [{\citenamefont {Pouiklis}\ \emph {et~al.}(2013)\citenamefont
  {Pouiklis}, \citenamefont {Kottaras}, \citenamefont {Psomoulis},\ and\
  \citenamefont {Sarris}}]{radiationhardened}%
  \BibitemOpen
  \bibfield  {author} {\bibinfo {author} {\bibfnamefont {G.}~\bibnamefont
  {Pouiklis}}, \bibinfo {author} {\bibfnamefont {G.}~\bibnamefont {Kottaras}},
  \bibinfo {author} {\bibfnamefont {A.}~\bibnamefont {Psomoulis}}, \ and\
  \bibinfo {author} {\bibfnamefont {E.}~\bibnamefont {Sarris}},\ }\href
  {\doibase 10.1080/00207217.2012.727349} {\bibfield  {journal} {\bibinfo
  {journal} {International Journal of Electronics}\ }\textbf {\bibinfo {volume}
  {100}},\ \bibinfo {pages} {913} (\bibinfo {year} {2013})}\BibitemShut
  {NoStop}%
\bibitem [{\citenamefont {Houck}\ \emph {et~al.}(2008)\citenamefont {Houck},
  \citenamefont {Schreier}, \citenamefont {Johnson}, \citenamefont {Chow},
  \citenamefont {Koch}, \citenamefont {Gambetta}, \citenamefont {Schuster},
  \citenamefont {Frunzio}, \citenamefont {Devoret}, \citenamefont {Girvin},\
  and\ \citenamefont {Schoelkopf}}]{coherencetime}%
  \BibitemOpen
  \bibfield  {author} {\bibinfo {author} {\bibfnamefont {A.~A.}\ \bibnamefont
  {Houck}}, \bibinfo {author} {\bibfnamefont {J.~A.}\ \bibnamefont {Schreier}},
  \bibinfo {author} {\bibfnamefont {B.~R.}\ \bibnamefont {Johnson}}, \bibinfo
  {author} {\bibfnamefont {J.~M.}\ \bibnamefont {Chow}}, \bibinfo {author}
  {\bibfnamefont {J.}~\bibnamefont {Koch}}, \bibinfo {author} {\bibfnamefont
  {J.~M.}\ \bibnamefont {Gambetta}}, \bibinfo {author} {\bibfnamefont {D.~I.}\
  \bibnamefont {Schuster}}, \bibinfo {author} {\bibfnamefont {L.}~\bibnamefont
  {Frunzio}}, \bibinfo {author} {\bibfnamefont {M.~H.}\ \bibnamefont
  {Devoret}}, \bibinfo {author} {\bibfnamefont {S.~M.}\ \bibnamefont {Girvin}},
  \ and\ \bibinfo {author} {\bibfnamefont {R.~J.}\ \bibnamefont {Schoelkopf}},\
  }\href {\doibase 10.1103/PhysRevLett.101.080502} {\bibfield  {journal}
  {\bibinfo  {journal} {Phys. Rev. Lett.}\ }\textbf {\bibinfo {volume} {101}},\
  \bibinfo {pages} {080502} (\bibinfo {year} {2008})}\BibitemShut {NoStop}%
\bibitem [{\citenamefont {Jeffrey}\ \emph {et~al.}(2014)\citenamefont
  {Jeffrey}, \citenamefont {Sank}, \citenamefont {Mutus}, \citenamefont
  {White}, \citenamefont {Kelly}, \citenamefont {Barends}, \citenamefont
  {Chen}, \citenamefont {Chen}, \citenamefont {Chiaro}, \citenamefont
  {Dunsworth}, \citenamefont {Megrant}, \citenamefont {O'Malley}, \citenamefont
  {Neill}, \citenamefont {Roushan}, \citenamefont {Vainsencher}, \citenamefont
  {Wenner}, \citenamefont {Cleland},\ and\ \citenamefont
  {Martinis}}]{HiFistateprep}%
  \BibitemOpen
  \bibfield  {author} {\bibinfo {author} {\bibfnamefont {E.}~\bibnamefont
  {Jeffrey}}, \bibinfo {author} {\bibfnamefont {D.}~\bibnamefont {Sank}},
  \bibinfo {author} {\bibfnamefont {J.~Y.}\ \bibnamefont {Mutus}}, \bibinfo
  {author} {\bibfnamefont {T.~C.}\ \bibnamefont {White}}, \bibinfo {author}
  {\bibfnamefont {J.}~\bibnamefont {Kelly}}, \bibinfo {author} {\bibfnamefont
  {R.}~\bibnamefont {Barends}}, \bibinfo {author} {\bibfnamefont
  {Y.}~\bibnamefont {Chen}}, \bibinfo {author} {\bibfnamefont {Z.}~\bibnamefont
  {Chen}}, \bibinfo {author} {\bibfnamefont {B.}~\bibnamefont {Chiaro}},
  \bibinfo {author} {\bibfnamefont {A.}~\bibnamefont {Dunsworth}}, \bibinfo
  {author} {\bibfnamefont {A.}~\bibnamefont {Megrant}}, \bibinfo {author}
  {\bibfnamefont {P.~J.~J.}\ \bibnamefont {O'Malley}}, \bibinfo {author}
  {\bibfnamefont {C.}~\bibnamefont {Neill}}, \bibinfo {author} {\bibfnamefont
  {P.}~\bibnamefont {Roushan}}, \bibinfo {author} {\bibfnamefont
  {A.}~\bibnamefont {Vainsencher}}, \bibinfo {author} {\bibfnamefont
  {J.}~\bibnamefont {Wenner}}, \bibinfo {author} {\bibfnamefont {A.~N.}\
  \bibnamefont {Cleland}}, \ and\ \bibinfo {author} {\bibfnamefont {J.~M.}\
  \bibnamefont {Martinis}},\ }\href {\doibase 10.1103/PhysRevLett.112.190504}
  {\bibfield  {journal} {\bibinfo  {journal} {Phys. Rev. Lett.}\ }\textbf
  {\bibinfo {volume} {112}},\ \bibinfo {pages} {190504} (\bibinfo {year}
  {2014})}\BibitemShut {NoStop}%
\bibitem [{\citenamefont {Plantenberg}\ \emph {et~al.}(2007)\citenamefont
  {Plantenberg}, \citenamefont {de~Groot}, \citenamefont {Harmans},\ and\
  \citenamefont {Mooij}}]{Qubitgates}%
  \BibitemOpen
  \bibfield  {author} {\bibinfo {author} {\bibfnamefont {J.}~\bibnamefont
  {Plantenberg}}, \bibinfo {author} {\bibfnamefont {P.}~\bibnamefont
  {de~Groot}}, \bibinfo {author} {\bibfnamefont {C.}~\bibnamefont {Harmans}}, \
  and\ \bibinfo {author} {\bibfnamefont {J.}~\bibnamefont {Mooij}},\ }\href
  {\doibase 10.1038/nature05896} {\bibfield  {journal} {\bibinfo  {journal}
  {Nature}\ }\textbf {\bibinfo {volume} {447}},\ \bibinfo {pages} {836}
  (\bibinfo {year} {2007})}\BibitemShut {NoStop}%
\bibitem [{\citenamefont {Steffen}\ \emph {et~al.}(2009)\citenamefont
  {Steffen}, \citenamefont {Brito}, \citenamefont {DiVincenzo}, \citenamefont
  {Kumar},\ and\ \citenamefont {Ketchen}}]{1367-2630-11-3-033030}%
  \BibitemOpen
  \bibfield  {author} {\bibinfo {author} {\bibfnamefont {M.}~\bibnamefont
  {Steffen}}, \bibinfo {author} {\bibfnamefont {F.}~\bibnamefont {Brito}},
  \bibinfo {author} {\bibfnamefont {D.}~\bibnamefont {DiVincenzo}}, \bibinfo
  {author} {\bibfnamefont {S.}~\bibnamefont {Kumar}}, \ and\ \bibinfo {author}
  {\bibfnamefont {M.}~\bibnamefont {Ketchen}},\ }\href
  {http://stacks.iop.org/1367-2630/11/i=3/a=033030} {\bibfield  {journal}
  {\bibinfo  {journal} {New Journal of Physics}\ }\textbf {\bibinfo {volume}
  {11}},\ \bibinfo {pages} {033030} (\bibinfo {year} {2009})}\BibitemShut
  {NoStop}%
\bibitem [{\citenamefont {Morshed}\ \emph {et~al.}(2011)\citenamefont {Morshed}
  \emph {et~al.}}]{MOSFET}%
  \BibitemOpen
  \bibfield  {author} {\bibinfo {author} {\bibfnamefont {T.~H.}\ \bibnamefont
  {Morshed}} \emph {et~al.},\ }\href@noop {} {\bibfield  {journal} {\bibinfo
  {journal} {BSIM4v4.7 MOSFET Model-User's Manual, University of California,
  Berkley}\ } (\bibinfo {year} {2011})}\BibitemShut {NoStop}%
\bibitem [{\citenamefont {Arrigoni}\ \emph {et~al.}(2013)\citenamefont
  {Arrigoni}, \citenamefont {Knap},\ and\ \citenamefont {von~der
  Linden}}]{PhysRevLett.110.086403}%
  \BibitemOpen
  \bibfield  {author} {\bibinfo {author} {\bibfnamefont {E.}~\bibnamefont
  {Arrigoni}}, \bibinfo {author} {\bibfnamefont {M.}~\bibnamefont {Knap}}, \
  and\ \bibinfo {author} {\bibfnamefont {W.}~\bibnamefont {von~der Linden}},\
  }\href {\doibase 10.1103/PhysRevLett.110.086403} {\bibfield  {journal}
  {\bibinfo  {journal} {Phys. Rev. Lett.}\ }\textbf {\bibinfo {volume} {110}},\
  \bibinfo {pages} {086403} (\bibinfo {year} {2013})}\BibitemShut {NoStop}%
\bibitem [{\citenamefont {Breuer}\ and\ \citenamefont
  {Petruccione}(2007)}]{breuer2007theory}%
  \BibitemOpen
  \bibfield  {author} {\bibinfo {author} {\bibfnamefont {H.}~\bibnamefont
  {Breuer}}\ and\ \bibinfo {author} {\bibfnamefont {F.}~\bibnamefont
  {Petruccione}},\ }\href {https://books.google.co.uk/books?id=DkcJPwAACAAJ}
  {\emph {\bibinfo {title} {The Theory of Open Quantum Systems}}}\ (\bibinfo
  {publisher} {OUP Oxford},\ \bibinfo {year} {2007})\BibitemShut {NoStop}%
\bibitem [{\citenamefont {Dyre}(1987)}]{PhysRevLett.58.792}%
  \BibitemOpen
  \bibfield  {author} {\bibinfo {author} {\bibfnamefont {J.~C.}\ \bibnamefont
  {Dyre}},\ }\href {\doibase 10.1103/PhysRevLett.58.792} {\bibfield  {journal}
  {\bibinfo  {journal} {Phys. Rev. Lett.}\ }\textbf {\bibinfo {volume} {58}},\
  \bibinfo {pages} {792} (\bibinfo {year} {1987})}\BibitemShut {NoStop}%
\bibitem [{\citenamefont {Giovannetti}\ and\ \citenamefont
  {Palma}(2012)}]{PhysRevLett.108.040401}%
  \BibitemOpen
  \bibfield  {author} {\bibinfo {author} {\bibfnamefont {V.}~\bibnamefont
  {Giovannetti}}\ and\ \bibinfo {author} {\bibfnamefont {G.~M.}\ \bibnamefont
  {Palma}},\ }\href {\doibase 10.1103/PhysRevLett.108.040401} {\bibfield
  {journal} {\bibinfo  {journal} {Phys. Rev. Lett.}\ }\textbf {\bibinfo
  {volume} {108}},\ \bibinfo {pages} {040401} (\bibinfo {year}
  {2012})}\BibitemShut {NoStop}%
\bibitem [{\citenamefont {Wu}\ \emph {et~al.}(2009)\citenamefont {Wu},
  \citenamefont {Kurizki},\ and\ \citenamefont
  {Brumer}}]{PhysRevLett.102.080405}%
  \BibitemOpen
  \bibfield  {author} {\bibinfo {author} {\bibfnamefont {L.-A.}\ \bibnamefont
  {Wu}}, \bibinfo {author} {\bibfnamefont {G.}~\bibnamefont {Kurizki}}, \ and\
  \bibinfo {author} {\bibfnamefont {P.}~\bibnamefont {Brumer}},\ }\href
  {\doibase 10.1103/PhysRevLett.102.080405} {\bibfield  {journal} {\bibinfo
  {journal} {Phys. Rev. Lett.}\ }\textbf {\bibinfo {volume} {102}},\ \bibinfo
  {pages} {080405} (\bibinfo {year} {2009})}\BibitemShut {NoStop}%
\bibitem [{\citenamefont {Ferialdi}(2016)}]{PhysRevLett.116.120402}%
  \BibitemOpen
  \bibfield  {author} {\bibinfo {author} {\bibfnamefont {L.}~\bibnamefont
  {Ferialdi}},\ }\href {\doibase 10.1103/PhysRevLett.116.120402} {\bibfield
  {journal} {\bibinfo  {journal} {Phys. Rev. Lett.}\ }\textbf {\bibinfo
  {volume} {116}},\ \bibinfo {pages} {120402} (\bibinfo {year}
  {2016})}\BibitemShut {NoStop}%
\bibitem [{\citenamefont {Hall}\ \emph {et~al.}(2014)\citenamefont {Hall},
  \citenamefont {Cresser}, \citenamefont {Li},\ and\ \citenamefont
  {Andersson}}]{PhysRevA.89.042120}%
  \BibitemOpen
  \bibfield  {author} {\bibinfo {author} {\bibfnamefont {M.~J.~W.}\
  \bibnamefont {Hall}}, \bibinfo {author} {\bibfnamefont {J.~D.}\ \bibnamefont
  {Cresser}}, \bibinfo {author} {\bibfnamefont {L.}~\bibnamefont {Li}}, \ and\
  \bibinfo {author} {\bibfnamefont {E.}~\bibnamefont {Andersson}},\ }\href
  {\doibase 10.1103/PhysRevA.89.042120} {\bibfield  {journal} {\bibinfo
  {journal} {Phys. Rev. A}\ }\textbf {\bibinfo {volume} {89}},\ \bibinfo
  {pages} {042120} (\bibinfo {year} {2014})}\BibitemShut {NoStop}%
\bibitem [{\citenamefont {Maldonado-Mundo}\ \emph {et~al.}(2012)\citenamefont
  {Maldonado-Mundo}, \citenamefont {\"Ohberg}, \citenamefont {Lovett},\ and\
  \citenamefont {Andersson}}]{PhysRevA.86.042107}%
  \BibitemOpen
  \bibfield  {author} {\bibinfo {author} {\bibfnamefont {D.}~\bibnamefont
  {Maldonado-Mundo}}, \bibinfo {author} {\bibfnamefont {P.}~\bibnamefont
  {\"Ohberg}}, \bibinfo {author} {\bibfnamefont {B.~W.}\ \bibnamefont
  {Lovett}}, \ and\ \bibinfo {author} {\bibfnamefont {E.}~\bibnamefont
  {Andersson}},\ }\href {\doibase 10.1103/PhysRevA.86.042107} {\bibfield
  {journal} {\bibinfo  {journal} {Phys. Rev. A}\ }\textbf {\bibinfo {volume}
  {86}},\ \bibinfo {pages} {042107} (\bibinfo {year} {2012})}\BibitemShut
  {NoStop}%
\bibitem [{\citenamefont {Joshi}\ \emph {et~al.}(2014)\citenamefont {Joshi},
  \citenamefont {\"Ohberg}, \citenamefont {Cresser},\ and\ \citenamefont
  {Andersson}}]{PhysRevA.90.063815}%
  \BibitemOpen
  \bibfield  {author} {\bibinfo {author} {\bibfnamefont {C.}~\bibnamefont
  {Joshi}}, \bibinfo {author} {\bibfnamefont {P.}~\bibnamefont {\"Ohberg}},
  \bibinfo {author} {\bibfnamefont {J.~D.}\ \bibnamefont {Cresser}}, \ and\
  \bibinfo {author} {\bibfnamefont {E.}~\bibnamefont {Andersson}},\ }\href
  {\doibase 10.1103/PhysRevA.90.063815} {\bibfield  {journal} {\bibinfo
  {journal} {Phys. Rev. A}\ }\textbf {\bibinfo {volume} {90}},\ \bibinfo
  {pages} {063815} (\bibinfo {year} {2014})}\BibitemShut {NoStop}%
\bibitem [{\citenamefont {Vacchini}(2016)}]{PhysRevLett.117.230401}%
  \BibitemOpen
  \bibfield  {author} {\bibinfo {author} {\bibfnamefont {B.}~\bibnamefont
  {Vacchini}},\ }\href {\doibase 10.1103/PhysRevLett.117.230401} {\bibfield
  {journal} {\bibinfo  {journal} {Phys. Rev. Lett.}\ }\textbf {\bibinfo
  {volume} {117}},\ \bibinfo {pages} {230401} (\bibinfo {year}
  {2016})}\BibitemShut {NoStop}%
\bibitem [{\citenamefont {Di\'osi}\ and\ \citenamefont
  {Ferialdi}(2014)}]{PhysRevLett.113.200403}%
  \BibitemOpen
  \bibfield  {author} {\bibinfo {author} {\bibfnamefont {L.}~\bibnamefont
  {Di\'osi}}\ and\ \bibinfo {author} {\bibfnamefont {L.}~\bibnamefont
  {Ferialdi}},\ }\href {\doibase 10.1103/PhysRevLett.113.200403} {\bibfield
  {journal} {\bibinfo  {journal} {Phys. Rev. Lett.}\ }\textbf {\bibinfo
  {volume} {113}},\ \bibinfo {pages} {200403} (\bibinfo {year}
  {2014})}\BibitemShut {NoStop}%
\bibitem [{\citenamefont {Lindblad}(1976)}]{LINDBLAD1976393}%
  \BibitemOpen
  \bibfield  {author} {\bibinfo {author} {\bibfnamefont {G.}~\bibnamefont
  {Lindblad}},\ }\href {\doibase
  http://dx.doi.org/10.1016/0034-4877(76)90029-X} {\bibfield  {journal}
  {\bibinfo  {journal} {Reports on Mathematical Physics}\ }\textbf {\bibinfo
  {volume} {10}},\ \bibinfo {pages} {393 } (\bibinfo {year}
  {1976})}\BibitemShut {NoStop}%
\bibitem [{\citenamefont {Caldeira}\ and\ \citenamefont
  {Leggett}(1983)}]{CALDEIRA1983374}%
  \BibitemOpen
  \bibfield  {author} {\bibinfo {author} {\bibfnamefont {A.}~\bibnamefont
  {Caldeira}}\ and\ \bibinfo {author} {\bibfnamefont {A.}~\bibnamefont
  {Leggett}},\ }\href {\doibase http://dx.doi.org/10.1016/0003-4916(83)90202-6}
  {\bibfield  {journal} {\bibinfo  {journal} {Annals of Physics}\ }\textbf
  {\bibinfo {volume} {149}},\ \bibinfo {pages} {374 } (\bibinfo {year}
  {1983})}\BibitemShut {NoStop}%
\bibitem [{\citenamefont {Gao}(1997)}]{PhysRevLett.79.3101}%
  \BibitemOpen
  \bibfield  {author} {\bibinfo {author} {\bibfnamefont {S.}~\bibnamefont
  {Gao}},\ }\href {\doibase 10.1103/PhysRevLett.79.3101} {\bibfield  {journal}
  {\bibinfo  {journal} {Phys. Rev. Lett.}\ }\textbf {\bibinfo {volume} {79}},\
  \bibinfo {pages} {3101} (\bibinfo {year} {1997})}\BibitemShut {NoStop}%
\bibitem [{\citenamefont {Sǎndulescu}\ and\ \citenamefont
  {Scutaru}(1987)}]{SANDULESCU1987277}%
  \BibitemOpen
  \bibfield  {author} {\bibinfo {author} {\bibfnamefont {A.}~\bibnamefont
  {Sǎndulescu}}\ and\ \bibinfo {author} {\bibfnamefont {H.}~\bibnamefont
  {Scutaru}},\ }\href {\doibase http://dx.doi.org/10.1016/0003-4916(87)90162-X}
  {\bibfield  {journal} {\bibinfo  {journal} {Annals of Physics}\ }\textbf
  {\bibinfo {volume} {173}},\ \bibinfo {pages} {277 } (\bibinfo {year}
  {1987})}\BibitemShut {NoStop}%
\bibitem [{\citenamefont {Wiseman}\ and\ \citenamefont
  {Munro}(1998)}]{PhysRevLett.80.5702}%
  \BibitemOpen
  \bibfield  {author} {\bibinfo {author} {\bibfnamefont {H.~M.}\ \bibnamefont
  {Wiseman}}\ and\ \bibinfo {author} {\bibfnamefont {W.~J.}\ \bibnamefont
  {Munro}},\ }\href {\doibase 10.1103/PhysRevLett.80.5702} {\bibfield
  {journal} {\bibinfo  {journal} {Phys. Rev. Lett.}\ }\textbf {\bibinfo
  {volume} {80}},\ \bibinfo {pages} {5702} (\bibinfo {year}
  {1998})}\BibitemShut {NoStop}%
\bibitem [{\citenamefont {Lampo}\ \emph {et~al.}(2016)\citenamefont {Lampo},
  \citenamefont {Lim}, \citenamefont {Wehr}, \citenamefont {Massignan},\ and\
  \citenamefont {Lewenstein}}]{PhysRevA.94.042123}%
  \BibitemOpen
  \bibfield  {author} {\bibinfo {author} {\bibfnamefont {A.}~\bibnamefont
  {Lampo}}, \bibinfo {author} {\bibfnamefont {S.~H.}\ \bibnamefont {Lim}},
  \bibinfo {author} {\bibfnamefont {J.}~\bibnamefont {Wehr}}, \bibinfo {author}
  {\bibfnamefont {P.}~\bibnamefont {Massignan}}, \ and\ \bibinfo {author}
  {\bibfnamefont {M.}~\bibnamefont {Lewenstein}},\ }\href {\doibase
  10.1103/PhysRevA.94.042123} {\bibfield  {journal} {\bibinfo  {journal} {Phys.
  Rev. A}\ }\textbf {\bibinfo {volume} {94}},\ \bibinfo {pages} {042123}
  (\bibinfo {year} {2016})}\BibitemShut {NoStop}%
\bibitem [{\citenamefont {Schack}\ \emph {et~al.}(1995)\citenamefont {Schack},
  \citenamefont {Brun},\ and\ \citenamefont {Percival}}]{0305-4470-28-18-028}%
  \BibitemOpen
  \bibfield  {author} {\bibinfo {author} {\bibfnamefont {R.}~\bibnamefont
  {Schack}}, \bibinfo {author} {\bibfnamefont {T.~A.}\ \bibnamefont {Brun}}, \
  and\ \bibinfo {author} {\bibfnamefont {I.~C.}\ \bibnamefont {Percival}},\
  }\href {http://stacks.iop.org/0305-4470/28/i=18/a=028} {\bibfield  {journal}
  {\bibinfo  {journal} {Journal of Physics A: Mathematical and General}\
  }\textbf {\bibinfo {volume} {28}},\ \bibinfo {pages} {5401} (\bibinfo {year}
  {1995})}\BibitemShut {NoStop}%
\bibitem [{\citenamefont {Everitt}(2009)}]{1367-2630-11-1-013014}%
  \BibitemOpen
  \bibfield  {author} {\bibinfo {author} {\bibfnamefont {M.~J.}\ \bibnamefont
  {Everitt}},\ }\href {http://stacks.iop.org/1367-2630/11/i=1/a=013014}
  {\bibfield  {journal} {\bibinfo  {journal} {New Journal of Physics}\ }\textbf
  {\bibinfo {volume} {11}},\ \bibinfo {pages} {013014} (\bibinfo {year}
  {2009})}\BibitemShut {NoStop}%
\bibitem [{\citenamefont {Everitt}\ \emph {et~al.}(2010)\citenamefont
  {Everitt}, \citenamefont {Munro},\ and\ \citenamefont
  {Spiller}}]{Everitt20102809}%
  \BibitemOpen
  \bibfield  {author} {\bibinfo {author} {\bibfnamefont {M.}~\bibnamefont
  {Everitt}}, \bibinfo {author} {\bibfnamefont {W.}~\bibnamefont {Munro}}, \
  and\ \bibinfo {author} {\bibfnamefont {T.}~\bibnamefont {Spiller}},\ }\href
  {\doibase http://dx.doi.org/10.1016/j.physleta.2010.05.006} {\bibfield
  {journal} {\bibinfo  {journal} {Physics Letters A}\ }\textbf {\bibinfo
  {volume} {374}},\ \bibinfo {pages} {2809 } (\bibinfo {year}
  {2010})}\BibitemShut {NoStop}%
\bibitem [{\citenamefont {Everitt}\ \emph {et~al.}(2011)\citenamefont
  {Everitt}, \citenamefont {Munro},\ and\ \citenamefont
  {Spiller}}]{1742-6596-306-1-012045}%
  \BibitemOpen
  \bibfield  {author} {\bibinfo {author} {\bibfnamefont {M.~J.}\ \bibnamefont
  {Everitt}}, \bibinfo {author} {\bibfnamefont {W.~J.}\ \bibnamefont {Munro}},
  \ and\ \bibinfo {author} {\bibfnamefont {T.~P.}\ \bibnamefont {Spiller}},\
  }\href {http://stacks.iop.org/1742-6596/306/i=1/a=012045} {\bibfield
  {journal} {\bibinfo  {journal} {Journal of Physics: Conference Series}\
  }\textbf {\bibinfo {volume} {306}},\ \bibinfo {pages} {012045} (\bibinfo
  {year} {2011})}\BibitemShut {NoStop}%
\bibitem [{\citenamefont {Everitt}\ \emph {et~al.}(2009)\citenamefont
  {Everitt}, \citenamefont {Munro},\ and\ \citenamefont
  {Spiller}}]{PhysRevA.79.032328}%
  \BibitemOpen
  \bibfield  {author} {\bibinfo {author} {\bibfnamefont {M.~J.}\ \bibnamefont
  {Everitt}}, \bibinfo {author} {\bibfnamefont {W.~J.}\ \bibnamefont {Munro}},
  \ and\ \bibinfo {author} {\bibfnamefont {T.~P.}\ \bibnamefont {Spiller}},\
  }\href {\doibase 10.1103/PhysRevA.79.032328} {\bibfield  {journal} {\bibinfo
  {journal} {Phys. Rev. A}\ }\textbf {\bibinfo {volume} {79}},\ \bibinfo
  {pages} {032328} (\bibinfo {year} {2009})}\BibitemShut {NoStop}%
\bibitem [{\citenamefont {Duffus}\ \emph {et~al.}(2016)\citenamefont {Duffus},
  \citenamefont {Bjergstrom}, \citenamefont {Dwyer}, \citenamefont {Samson},
  \citenamefont {Spiller}, \citenamefont {Zagoskin}, \citenamefont {Munro},
  \citenamefont {Nemoto},\ and\ \citenamefont {Everitt}}]{PhysRevB.94.064518}%
  \BibitemOpen
  \bibfield  {author} {\bibinfo {author} {\bibfnamefont {S.~N.~A.}\
  \bibnamefont {Duffus}}, \bibinfo {author} {\bibfnamefont {K.~N.}\
  \bibnamefont {Bjergstrom}}, \bibinfo {author} {\bibfnamefont {V.~M.}\
  \bibnamefont {Dwyer}}, \bibinfo {author} {\bibfnamefont {J.~H.}\ \bibnamefont
  {Samson}}, \bibinfo {author} {\bibfnamefont {T.~P.}\ \bibnamefont {Spiller}},
  \bibinfo {author} {\bibfnamefont {A.~M.}\ \bibnamefont {Zagoskin}}, \bibinfo
  {author} {\bibfnamefont {W.~J.}\ \bibnamefont {Munro}}, \bibinfo {author}
  {\bibfnamefont {K.}~\bibnamefont {Nemoto}}, \ and\ \bibinfo {author}
  {\bibfnamefont {M.~J.}\ \bibnamefont {Everitt}},\ }\href {\doibase
  10.1103/PhysRevB.94.064518} {\bibfield  {journal} {\bibinfo  {journal} {Phys.
  Rev. B}\ }\textbf {\bibinfo {volume} {94}},\ \bibinfo {pages} {064518}
  (\bibinfo {year} {2016})}\BibitemShut {NoStop}%
\bibitem [{\citenamefont {Cole}\ \emph {et~al.}(2005)\citenamefont {Cole},
  \citenamefont {Schirmer}, \citenamefont {Greentree}, \citenamefont {Wellard},
  \citenamefont {Oi},\ and\ \citenamefont {Hollenberg}}]{PhysRevA.71.062312}%
  \BibitemOpen
  \bibfield  {author} {\bibinfo {author} {\bibfnamefont {J.~H.}\ \bibnamefont
  {Cole}}, \bibinfo {author} {\bibfnamefont {S.~G.}\ \bibnamefont {Schirmer}},
  \bibinfo {author} {\bibfnamefont {A.~D.}\ \bibnamefont {Greentree}}, \bibinfo
  {author} {\bibfnamefont {C.~J.}\ \bibnamefont {Wellard}}, \bibinfo {author}
  {\bibfnamefont {D.~K.~L.}\ \bibnamefont {Oi}}, \ and\ \bibinfo {author}
  {\bibfnamefont {L.~C.~L.}\ \bibnamefont {Hollenberg}},\ }\href {\doibase
  10.1103/PhysRevA.71.062312} {\bibfield  {journal} {\bibinfo  {journal} {Phys.
  Rev. A}\ }\textbf {\bibinfo {volume} {71}},\ \bibinfo {pages} {062312}
  (\bibinfo {year} {2005})}\BibitemShut {NoStop}%
\bibitem [{\citenamefont {Devitt}\ \emph {et~al.}(2013)\citenamefont {Devitt},
  \citenamefont {Munro},\ and\ \citenamefont {Nemoto}}]{0034-4885-76-7-076001}%
  \BibitemOpen
  \bibfield  {author} {\bibinfo {author} {\bibfnamefont {S.~J.}\ \bibnamefont
  {Devitt}}, \bibinfo {author} {\bibfnamefont {W.~J.}\ \bibnamefont {Munro}}, \
  and\ \bibinfo {author} {\bibfnamefont {K.}~\bibnamefont {Nemoto}},\ }\href
  {http://stacks.iop.org/0034-4885/76/i=7/a=076001} {\bibfield  {journal}
  {\bibinfo  {journal} {Reports on Progress in Physics}\ }\textbf {\bibinfo
  {volume} {76}},\ \bibinfo {pages} {076001} (\bibinfo {year}
  {2013})}\BibitemShut {NoStop}%
\bibitem [{\citenamefont {Diggins}\ \emph {et~al.}(1994)\citenamefont
  {Diggins}, \citenamefont {Ralph}, \citenamefont {Spiller}, \citenamefont
  {Clark}, \citenamefont {Prance},\ and\ \citenamefont
  {Prance}}]{PhysRevE.49.1854}%
  \BibitemOpen
  \bibfield  {author} {\bibinfo {author} {\bibfnamefont {J.}~\bibnamefont
  {Diggins}}, \bibinfo {author} {\bibfnamefont {J.~F.}\ \bibnamefont {Ralph}},
  \bibinfo {author} {\bibfnamefont {T.~P.}\ \bibnamefont {Spiller}}, \bibinfo
  {author} {\bibfnamefont {T.~D.}\ \bibnamefont {Clark}}, \bibinfo {author}
  {\bibfnamefont {H.}~\bibnamefont {Prance}}, \ and\ \bibinfo {author}
  {\bibfnamefont {R.~J.}\ \bibnamefont {Prance}},\ }\href {\doibase
  10.1103/PhysRevE.49.1854} {\bibfield  {journal} {\bibinfo  {journal} {Phys.
  Rev. E}\ }\textbf {\bibinfo {volume} {49}},\ \bibinfo {pages} {1854}
  (\bibinfo {year} {1994})}\BibitemShut {NoStop}%
\bibitem [{\citenamefont {Spiller}\ \emph {et~al.}(1991)\citenamefont
  {Spiller}, \citenamefont {Poulton}, \citenamefont {Clark}, \citenamefont
  {Prance},\ and\ \citenamefont {Prance}}]{Spiller1991}%
  \BibitemOpen
  \bibfield  {author} {\bibinfo {author} {\bibfnamefont {T.~P.}\ \bibnamefont
  {Spiller}}, \bibinfo {author} {\bibfnamefont {D.~A.}\ \bibnamefont
  {Poulton}}, \bibinfo {author} {\bibfnamefont {T.~D.}\ \bibnamefont {Clark}},
  \bibinfo {author} {\bibfnamefont {R.~J.}\ \bibnamefont {Prance}}, \ and\
  \bibinfo {author} {\bibfnamefont {H.}~\bibnamefont {Prance}},\ }\href
  {\doibase 10.1017/CBO9781107415324.004} {\bibfield  {journal} {\bibinfo
  {journal} {International Journal of Modern Physics B}\ }\textbf {\bibinfo
  {volume} {5}},\ \bibinfo {pages} {1437} (\bibinfo {year} {1991})},\ \Eprint
  {http://arxiv.org/abs/arXiv:1011.1669v3} {arXiv:arXiv:1011.1669v3}
  \BibitemShut {NoStop}%
\bibitem [{\citenamefont {Munro}\ and\ \citenamefont
  {Gardiner}(1996)}]{Munro1996}%
  \BibitemOpen
  \bibfield  {author} {\bibinfo {author} {\bibfnamefont {W.}~\bibnamefont
  {Munro}}\ and\ \bibinfo {author} {\bibfnamefont {C.}~\bibnamefont
  {Gardiner}},\ }\href {\doibase 10.1103/PhysRevA.53.2633} {\bibfield
  {journal} {\bibinfo  {journal} {Physical Review A}\ }\textbf {\bibinfo
  {volume} {53}},\ \bibinfo {pages} {2633} (\bibinfo {year}
  {1996})}\BibitemShut {NoStop}%
\bibitem [{\citenamefont {Prance}\ \emph {et~al.}(1985)\citenamefont {Prance},
  \citenamefont {Mutton}, \citenamefont {Shephard}, \citenamefont {Clark},
  \citenamefont {Prance},\ and\ \citenamefont {Spiller}}]{SQUID}%
  \BibitemOpen
  \bibfield  {author} {\bibinfo {author} {\bibfnamefont {R.~J.}\ \bibnamefont
  {Prance}}, \bibinfo {author} {\bibfnamefont {J.~E.}\ \bibnamefont {Mutton}},
  \bibinfo {author} {\bibfnamefont {E.~P.}\ \bibnamefont {Shephard}}, \bibinfo
  {author} {\bibfnamefont {T.~D.}\ \bibnamefont {Clark}}, \bibinfo {author}
  {\bibfnamefont {H.}~\bibnamefont {Prance}}, \ and\ \bibinfo {author}
  {\bibfnamefont {T.~P.}\ \bibnamefont {Spiller}},\ }\href@noop {} {\emph
  {\bibinfo {title} {SQUID '85 Superconducting Quantum Interference Devices and
  their Applications}}},\ edited by\ \bibinfo {editor} {\bibfnamefont {H.~D.}\
  \bibnamefont {Hahlbohm}}\ and\ \bibinfo {editor} {\bibfnamefont
  {H.}~\bibnamefont {Lubbig}}\ (\bibinfo {year} {1985})\BibitemShut {NoStop}%
\end{thebibliography}%

%
%
%
 

%
%
%
%

 \end{document}